\documentclass[a4paper,12pt]{article}
\pdfoutput=1
\usepackage{feynmp-auto,expdlist}
\usepackage{amsmath, amsfonts, amssymb}
\usepackage{graphicx}
\usepackage{enumerate}
\usepackage{hyperref}
\usepackage{latexsym}
\usepackage{dsfont}
\usepackage{hepnicenames}
\usepackage{enumerate}
\usepackage{soul}
\usepackage[normalem]{ulem}
\usepackage{comment}

\usepackage{units}

\usepackage{mathrsfs,graphicx,rotating,amsmath,amsfonts,mathtools,booktabs,amssymb,wasysym}
\usepackage{hyperref}\usepackage{slashed}
\usepackage[nosort]{cite}
\usepackage[table,xcdraw,dvipsnames]{xcolor}
\usepackage{bm}
\usepackage{graphicx}
\usepackage{multirow,multicol}
\hypersetup{colorlinks,bookmarksopen,bookmarksnumbered,
	linkcolor=blus,pdfstartview=FitH,urlcolor=rossos,citecolor=verde}
\allowdisplaybreaks

\renewcommand{\[}{\left[}

\newcommand{\cW}{c_{\rm W}}

\def\Lag{\mathscr{L}}

\newcommand{\mio}[1]{}
\newcommand{\BR}{\hbox{BR}}

\def\bpm{\begin{pmatrix}}
	\def\epm{\end{pmatrix}}

\usepackage{mathrsfs}

\newcommand{\fig}[1]{~\ref{fig:#1}}

\allowdisplaybreaks
\usepackage{multicol}
\usepackage{color}
\definecolor{rosso}{cmyk}{0,1,1,0.4}
\definecolor{rossos}{cmyk}{0,1,1,0.55}
\definecolor{rossoc}{cmyk}{0,1,1,0.2}
\definecolor{blu}{cmyk}{1,1,0,0.3}
\definecolor{blus}{cmyk}{1,1,0,0.6}
\definecolor{bluc}{cmyk}{1,1,0,0.1}
\definecolor{verde}{cmyk}{0.92,0,0.59,0.25}
\definecolor{verdec}{cmyk}{0.92,0,0.59,0.15}
\definecolor{verdes}{cmyk}{0.92,0,0.59,0.4}

\newcommand{\eq}[1]{~{\rm (\ref{eq:#1})}}

\newcommand{\keV}{\,{\rm keV}}

\newcommand{\MeV}{\,{\rm MeV}}
\newcommand{\GeV}{\,{\rm GeV}}
\newcommand{\TeV}{\,{\rm TeV}}
\newcommand{\cm}{\,{\rm cm}}
\newcommand{\fb}{\,{\rm fb}}
\newcommand{\ab}{\,{\rm ab}}

\def\circa#1{\,\raise.3ex\hbox{$#1$\kern-.75em\lower1ex\hbox{$\sim$}}\,}

\newcommand{\beq}{\begin{equation}}
\newcommand{\eeq}{\end{equation}}

\newcommand{\bea}{\begin{eqnarray}}
\newcommand{\eea}{\end{eqnarray}}
\newcommand{\be}{\begin{equation}}
\newcommand{\ee}{\end{equation}}
\font\tenrsfs=rsfs10 at 12pt
\font\sevenrsfs=rsfs7
\font\fiversfs=rsfs5
\newfam\rsfsfam
\textfont\rsfsfam=\tenrsfs
\scriptfont\rsfsfam=\sevenrsfs
\scriptscriptfont\rsfsfam=\fiversfs

\newsavebox\MBox

\renewenvironment{thebibliography}[1]
{\begin{multicols}{2}[\section*{\refname}]%
		\@mkboth{\MakeUppercase\refname}{\MakeUppercase\refname}%
		\list{\@biblabel{\@arabic\c@enumiv}}%
		{\settowidth\labelwidth{\@biblabel{#1}}%
			\leftmargin\labelwidth
			\advance\leftmargin\labelsep
			\@openbib@code
			\usecounter{enumiv}%
			\let\p@enumiv\@empty
			\renewcommand\theenumiv{\@arabic\c@enumiv}}%
		\sloppy
		\clubpenalty4000
		\@clubpenalty \clubpenalty
		\widowpenalty4000%
		\sfcode`\.\@m}
	{\def\@noitemerr
		{\@latex@warning{Empty `thebibliography' environment}}%
		\endlist\end{multicols}}

\newcommand{\alf}{\alpha_{\rm eff}}
\newcommand{\eV}{\,{\rm eV}}
\newcommand{\SU}{\,{\rm SU}}

\def\circa#1{\,\raise.3ex\hbox{$#1$\kern-.75em\lower1ex\hbox{$\sim$}}\,}
\makeatletter

\font\ital=cmu10

\def\hhref#1{\href{http://arxiv.org/abs/#1}{arXiv:#1}}
\usepackage{xstring}
\newcommand{\hhrefq}[1]{\IfSubStr{#1}{:}{\href{http://inspirehep.net/search?ln=en&ln=en&p=#1&of=hb&action_search=Search&sf=&so=d&rm=&rg=25&sc=0}{InSpire:#1}}{\hhref{#1}}}

\def\art{\@ifnextchar[{\eart}{\oart}}
\def\eart[#1]#2#3#4#5#6{{\rm #2}, {\em #3 \bf #4} {\rm (#6) #5} ({\em #1})}
\def\article{\@ifnextchar[{\earticle}{\oarticle}}
\def\oarticle#1#2#3#4#5#6{{\rm #1}, {\ital ``#6''}, {\rm #2 #3 (#5) #4}}
\def\earticle[#1]#2#3#4#5#6#7{{\rm #2}, {\ital ``#7''}, {\rm #3 #4 (#6) #5}  [\hhrefq{#1}]}
\def\hepart[#1]#2{{\rm #2, \sl#1}}
\def\heparticle[#1]#2#3{#2, {\ital ``#3''} [\hhrefq{#1}]}
\newcommand{\doi}[1]{\href{http://dx.doi.org/#1}{[link]}}

\newcommand{\hhrefqq}[1]{\IfBeginWith{#1}{10.}{\href{https://doi.org/#1}{doi:#1}}{\hhrefq{#1}}}
\def\earticle[#1]#2#3#4#5#6#7{{\rm #2}, {\ital ``#7''}, {\rm #3 #4 (#6) #5}  [\hhrefqq{#1}]}

\renewenvironment{thebibliography}[1]
{\begin{multicols}{2}[\section*{\refname}]%
		\@mkboth{\MakeUppercase\refname}{\MakeUppercase\refname}%
		\list{\@biblabel{\@arabic\c@enumiv}}%
		{\settowidth\labelwidth{\@biblabel{#1}}%
			\leftmargin\labelwidth
			\advance\leftmargin\labelsep
			\@openbib@code
			\usecounter{enumiv}%
			\let\p@enumiv\@empty
			\renewcommand\theenumiv{\@arabic\c@enumiv}}%
		\sloppy
		\clubpenalty4000
		\@clubpenalty \clubpenalty
		\widowpenalty4000%
		\sfcode`\.\@m}
	{\def\@noitemerr
		{\@latex@warning{Empty `thebibliography' environment}}%
		\endlist\end{multicols}}

\newcounter{alphaequation}[equation]
\def\thealphaequation{\theequation\hbox to
	0.6em{\hfil\alph{alphaequation}\hfil}}
\def\eqnsystem#1{
	\def\@eqnnum{{\rm (\thealphaequation)}}
	\def\@@eqncr{\let\@tempa\relax \ifcase\@eqcnt \def\@tempa{& & &} \or
		\def\@tempa{& &}\or \def\@tempa{&}\fi\@tempa
		\if@eqnsw\@eqnnum\refstepcounter{alphaequation}\fi
		\global\@eqnswtrue\global\@eqcnt=0\cr}
	\refstepcounter{equation} \let\@currentlabel\theequation \def\@tempb{#1}
	\ifx\@tempb\empty\else\label{#1}\fi
	\refstepcounter{alphaequation}
	\let\@currentlabel\thealphaequation
	\global\@eqnswtrue\global\@eqcnt=0 \tabskip\@centering\let\\=\@eqncr
	$$\halign to \displaywidth\bgroup \@eqnsel\hskip\@centering
	$\displaystyle\tabskip\z@{##}$&\global\@eqcnt\@ne
	\hskip2\arraycolsep\hfil${##}$\hfil& \global\@eqcnt\tw@\hskip2\arraycolsep
	$\displaystyle\tabskip\z@{##}$\hfil
	\tabskip\@centering&\llap{##}\tabskip\z@\cr}
\def\endeqnsystem{\@@eqncr\egroup$$\global\@ignoretrue} \makeatother

\definecolor{Gray}{gray}{0.95}

\def\bal#1\eal{\begin{align}#1\end{align}}

\begin{document}
\thispagestyle{empty}

\begin{center}  
{\Large\Huge\bf\color{rossos} Minimal Dark Matter bound\\[1ex] states at future colliders} \\
\vspace{1cm}
{\bf  Salvatore Bottaro$^a$, Alessandro Strumia$^{b}$, Natascia Vignaroli$^b$}\\[7mm]

{\it $^a$ Scuola Normale, Pisa, Italia}\\[1mm]
{\it $^b$ Dipartimento di Fisica, Universit\`a di Pisa, Italia}\\[1mm]

\vspace{1cm}

{\large\bf Abstract}
\begin{quote}\large
The hypothesis that Dark Matter is one electroweak multiplet
leads to predictive candidates with multi-TeV masses that can form electroweak bound states.
Bound states with the same quantum numbers as electroweak vectors
are found to be especially interesting, 
as they can be produced resonantly with large cross sections at lepton colliders.
Such bound states exist e.g.\ if  DM is an automatically stable fermionic weak 5-plet
with mass $M \approx 14\TeV$ such that the DM abundance is reproduced thermally.
In this model, a muon collider could resolve three such bound states.
Production rates are so large that details of DM spectroscopy
can be probed with larger statistics:
we compute the characteristic pattern of single and multiple $\gamma$ lines.
\end{quote}
\end{center}

\setcounter{page}{1}

\vfill\vfill\tableofcontents\vfill\newpage
	
\section{Introduction}
The hypothesis that Dark Matter is the thermal relic of one new multiplet under the Standard Model gauge group
provides some predictive allowed candidates~\cite{hep-ph/0512090}.
In particular, the cosmological DM abundance is reproduced for TeV-scale values of the DM mass,
above the LHC reach.
A fermionic 5-plet under $\SU(2)_L$ with zero hypercharge is a particularly interesting possibility,
being automatically long lived enough to be DM.
Its thermal abundance matches the DM density for a mass $M \approx 14\TeV$,
after taking into account Sommerfeld and bound-state corrections~\cite{1702.01141}.
Its univocal prediction for direct detection can be tested~\cite{hep-ph/0512090,1504.00915},
but its production at colliders would allow to measure more than one number.
However, even a giant $pp$ collider at 100 TeV would have a limited reach, up to about 4 TeV~\cite{100TeV}
(see also~\cite{1407.7058,1506.03445,1805.00015,1901.02987}).

\medskip

A future muon collider could be built
in the existing 27 km LEP/LHC circular tunnel.
In such a case, $\mu^\pm$ beams  can reach the maximal energy $\sqrt{s}$ allowed by magnetic fields.
This was $\sqrt{s}\approx 14\TeV$ at LHC, but future magnets can realistically increase it up to $\sqrt{s}\sim 30\TeV$.
Concerning the integrated luminosity, a value ${\cal L} \sim 90/\ab$ at $\sqrt{s}=30\TeV$ is considered possible~\cite{2009.11287},
provided that radiological hazards due to muon decays into neutrinos can be limited.

According to~\cite{2009.11287}, 
a muon collider with this luminosity cannot probe a Minimal DM 5-plet with $M \approx 14\TeV$,
unless experiments are able to find in the beam-related background
the tracks left by its  charged components, short because produced with non-relativistic velocity
(see~\cite{2102.11292} for possible strategies).
Otherwise, missing energy signals tagged by an extra muon or gamma
have low cross sections and can only probe 5-plets lighter than about 10 TeV~\cite{2009.11287}.

\medskip

We show that extra signals arise taking into account that such DM forms weak bound states
with binding energy $E_B \sim 100\GeV$.
Such bound states annihilate into SM particles 
(including $\mu^+\mu^-$, for appropriate bound states with the same quantum numbers
of electroweak vectors)
with a width, $\Gamma_B \sim \alpha_2^5 M$, that is small but not much
smaller than the expected energy resolution of a muon collider, $\sigma_E \sim 10^{-3} E$.
Production and annihilation of DM bound states $B$ thereby results into 
a large cross section among visible SM particles,
\beq \sigma (\mu^+\mu^-\to B \to f\bar f)\sim \sigma_{\rm peak} \frac{\Gamma_B}{\sigma_E} \qquad\hbox{
where}\qquad \sigma_{\rm peak}\sim \frac{4\pi}{s}\eeq 
is the maximal cross section allowed by unitarity.

One needs to run around the peak, and
the minimal 5-plet DM model allows to predict the DM mass, $M \approx 14\TeV$, from the cosmological DM density. 
More in general (for example a family of three 5-plets allows for smaller $M$)
one could first discover DM through other signals, measure its mass, and next run on the peak,
possibly gradually reducing the beam energy spread $\sigma_E$ of the collider
to achieve the maximal cross section.
The energy resolution of a muon collider can be reduced by at least one order of magnitude,
down to $\sigma_E \sim 10^{-4}E$, at the price of proportionally reducing its luminosity.
We also consider other bound states and $pp$ colliders, obtaining small cross sections as no resonant production is possible.\footnote{DM bound states of an electroweak triplet
have been discussed in~\cite{Katayose} at a $pp$ collider, where no resonant production is possible.
We here include important non-abelian Coulomb-like potentials.
See also~\cite{DMboundstatecollider}.}

The paper is structured as follows.
In section~\ref{MDM5} we summarize the DM model and the properties of DM bound states.
In section~\ref{boundcoll} we discuss bound state production at colliders,
focusing in~\ref{ns3} on the main signals from states that can be produced resonantly with large cross section, in~\ref{boundcollnonres}  on other bound states,
in~\ref{boundcolldec} on rarer but very characteristic signals coming from decays among DM bound states,
such as $\gamma$ lines.
In section~\ref{concl} we give conclusions, and mention one more (curious but small) signal of Minimal DM.

\begin{figure}[t]
\centering
$$\includegraphics[width=0.5\textwidth]{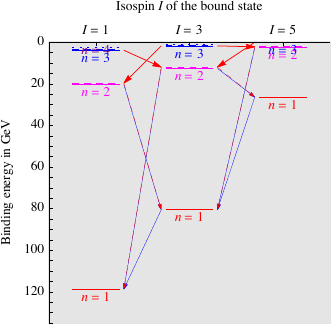}$$
\caption{\em\label{fig:Levels14TeV}
Energy levels of bound states of two Minimal Dark Matter 5-plets with $M=14\TeV$.
Continuous lines have $\ell=0$, dashed lines have $\ell=1$, dotted lines have $\ell=2$.
The blue (red) arrows indicate some main magnetic (electric) decays.
}
\end{figure}

\section{Minimal Dark Matter and its bound states}\label{MDM5}
The SM is extended adding a fermionic 5-plet ${\cal X}$ under $\SU(2)_L$ with zero hypercharge, such that the most general renormalizable Lagrangian is
\beq\Lag = \Lag_{\rm SM} +\frac12
 \bar{{\cal X}} (i  \slashed{D}+M) {\cal X} .\eeq
The 5-plet contains a Majorana neutral component ${\cal X}^0$, 
together with Dirac fermions ${\cal X}^\pm$ and ${\cal X}^{\pm\pm}$ with charge $\pm1$ and $\pm2$.
Due to its electric field, the components with charge $q$ are heavier than ${\cal X}^0$ by 
$q^2 \Delta M$, where $ \Delta M = \alpha_2 M_W \sin^2({\theta_{\rm W}}/{2}) \approx 166\MeV$.
Thereby charged states decay into the lightest DM-candidate ${\cal X}^0$. 

\smallskip

\begin{table}[t]
$$\resizebox{\columnwidth}{!}{\begin{tabular}{c|ccccc|rlc|cc}
\rowcolor[HTML]{D0D0D0} 
name &  \multicolumn{5}{c|}{Quantum numbers}   &  \multicolumn{3}{c|}{Annihilation}  &  \multicolumn{2}{c}{Decay}\\  
\rowcolor[HTML]{D0D0D0} 
\hbox{$^n_J\ell^{PC}_I$}& $n$ & $I$ & $S$  & $\ell$ & $E_B$ &  \multicolumn{2}{c}{$\Gamma_{\rm ann}$}  & \hbox{into} & $\Gamma_{\rm dec}$ & \hbox{into} \\  \toprule
\rowcolor[cmyk]{0,0,0.1,0}
$^1_1s_1^{-+}$ & 1 &$1$ & 0  & 0 & 118 GeV & $3240\,\alpha_2^5 M$ & $\!\! \approx 1.63\GeV$ & $V\tilde V$ & 0 &---   \\  \rowcolor[cmyk]{0,0,0.1,0}
$^1_1s_3^{--}$ & 1 & 3 & 1  & 0 & 81 GeV &   $15625 \,\alpha_2^5 M/48$ & $\!\! \approx 0.17\GeV$ & $f_L\bar f_L+HH^*$&$36\,\alpha_2^6\alpha_{\rm em} M\approx 4.6\keV$&$^1s_1 \gamma$ \\  \rowcolor[cmyk]{0,0,0.1,0}
$^1_1s_5^{-+}$ & 1 & 5 & 0  & 0  & 26 GeV &  $567\,\alpha_2^5 M /4 $ & $\!\! \approx0.07\GeV$ & $V\tilde V$ &$ 295\,\alpha_2^6\alpha_{\rm em} M\approx 38\keV$ & $^1s_3 \gamma$  \\  \midrule \rowcolor[cmyk]{0,0,0.2,0}
$^2_1s_1^{-+}$ & 2&1 & 0  & 0 & 20.3 GeV &   $405\alpha_2^5M$ & $\!\! \approx0.2\GeV$ & $V\tilde V$ & $13 \,\alpha_2^6  \alpha_{\rm em}M\approx 1.7\keV$ & $^1s_{3}\gamma$  \\ \rowcolor[cmyk]{0,0,0.2,0}
$^2_1s_3^{--}$ & 2& 3 & 1  & 0 & 13 GeV &  $15625 \, \alpha_2^5 M/384$ & $\!\! \approx21\MeV$  & $f_L \bar f_L+HH^*$ & $(6.9\,\alpha_2 + 0.3\,\alpha_{\rm em}) \alpha_2^6 M\approx 3.7\keV$ & $^1s_{1+5}V$  \\ \rowcolor[cmyk]{0,0,0.2,0}
$^2_1s_5^{-+}$ & 2 & 5 & 0 & 0  & 2.6 GeV &  $567\,\alpha_2^5 M /32$ & $\!\! \approx 9\MeV$ & $V\tilde V$ & $ 28.4\, \alpha_2^6  \alpha_{\rm em} M\approx 3.6 \keV$  & $^1s_{3}\gamma$ \\ \midrule  \rowcolor[cmyk]{0.1,0,0.1,0}
$^2_Jp_1^{++}$ & 2& 1 & 1  & 1  & 19.7 GeV  & ${\cal O}( \alpha_2^7M)$ & $\!\!\sim\keV$ & $VV$ & $ 20.4\,\alpha_2^4 \alpha_{\rm em}M \approx 2.5\MeV$  & $^1s_3 \gamma$  \\ \rowcolor[cmyk]{0.1,0,0.1,0}
$^2_1p_3^{+-}$ & 2& 3 & 0 & 1  & 12 GeV & ${\cal O}( \alpha_2^8M)$ & $\!\!\sim10\eV$  & $VVV$ &$(30.2\,\alpha_2 + 0.3\,\alpha_{\rm em})\, \alpha_2^4 M\approx 15.3\MeV $& $^1s_{1+5} V$ \\ \rowcolor[cmyk]{0.1,0,0.1,0}
$^2_Jp_5^{++}$ & 2& 5 & 1  & 1  &  2.2 GeV &  ${\cal O}( \alpha_2^7M)$ & $\!\!\sim\keV$ & $VV$ &$ 4.7\,\alpha_2^4 \alpha_{\rm em}M\approx 0.6\MeV$ & $^1s_{3} \gamma$ \\ \hline \rowcolor[cmyk]{0,0,0.3,0}
$^3_1s_1^{-+}$ & 3 & 1 & 0  & 0 & 3.8 GeV &  $120\alpha_2^5 M$ & $\!\! \approx 60\MeV$ & $V\tilde V$& $ 0.34\,\alpha_2^4\alpha_{\rm em}M \approx 42\keV $ & $^2p_{3} \gamma$\\  \rowcolor[cmyk]{0,0,0.3,0}
$^3_1s_3^{--}$ & 3 & 3 & 1  & 0 &  1.7 GeV  & $15625\alpha_2^5 M/1296$ & $\!\! \approx 6.0\MeV$ & $f_L\bar f_L+HH^*$& $(0.003+0.005)\alpha_2^4\alpha_{\rm em}M \approx 1\keV $ & $^2p_{1+5} \gamma$\\  \rowcolor[cmyk]{0,0,0.3,0}
$^3_1s_5^{-+}$ & 3 & 5 & 0  & 0 &   1.7 MeV & $21\alpha_2^5 M/4$ & $\!\! \approx 2.7\MeV$ & $V \tilde V$& $ 0.3\alpha_2^4\alpha_{\rm em}M \approx 36\keV $ & $^2p_{3} \gamma$\\ \hline  \rowcolor[cmyk]{0,0.2,0.2,0}
$^3_Jd_3^{--}$ & 3 & 3 & 1  & 2 &  0.9 GeV  & ${\cal O}( \alpha_2^9 M )$ & $\!\! \sim \eV$ & $f_L\bar f_L$& $0.4\alpha_2^4\alpha_{\rm em}M \approx 52\keV $ & $^{2}p_{1+5} \gamma$\\
\end{tabular}}$$
\caption{\label{tab:MainLevels}\em Main bound states of fermion weak 5-plets with $M \approx 14\TeV$. 
The parity $P=(-1)^{\ell +1}$ and charge conjugation $C= (-1)^{\ell+S}$
quantum numbers of bound states
are broken by chiral weak gauge interactions to SM fermions. 
Hyper-fine components with different values of $J$ have the same decay rate.
Decay rates are not $\SU(2)_L$-invariant because $W,Z$ emission is sometimes blocked by phase space;
we report decay rates averaged over the weak components of bound states.}
\end{table}%

The components of ${\cal X}$ can be pair-produced at collider.
Pairs of 5-plets can form Coulombian-like electroweak bound states given that $M \circa{>} M_{W,Z}/\alpha_2$. 
Bound states can be computed in components~\cite{0706.4071}, 
or more simply in $\SU(2)_L$-symmetric approximation, 
setting $M_W \approx M_Z$ and neglecting $\Delta M$~\cite{1702.01141}.
States of two quintuplets decompose under $\SU(2)_L$ as
$5 \otimes 5= 1_S \oplus 3_A \oplus 5_S \oplus 7_A \oplus 9_S$.
The attractive channels with potential $V = -\alpha_{\rm eff} e^{-M_{W,Z} r}/r$ have isospin
$I=1$ ($\alpha_{\rm eff}=6\alpha_2$),
$I=3$ ($\alpha_{\rm eff}=5\alpha_2$),
and $I=5$ ($\alpha_{\rm eff}=3\alpha_2$).
The SM weak couplings renormalized at TeV energy are
$\alpha_2 = 1/30.8$ and $\alpha_{\rm em} =1/128.6$.
The binding energies of bound states can be approximated as~\cite{1702.01141}
\beq\label{eq:EboundYuk}
E_B \approx  \frac{\alpha_\text{eff}^2 M}{4 n^2}  
 \bigg[  1-   n^2 y-0.53 n^2 y^2 \ell (\ell+1) \bigg]^2\qquad\hbox{where}\qquad
 y\approx \frac{1.74 M_{W,Z}}{\alf M} 
\eeq
and $\ell$ is angular momentum.
The bound state exists only when the term in the squared parenthesis is positive.
Furthermore, only bound states 
with $(-1)^{\ell +S +\tilde I}=1$ have the correct fermionic anti-symmetry,
where $I = 2 \tilde I+1$ is the dimension of the representation.
As usual for bound states of two fermions, our states
have quantum numbers $C=(-1)^{\ell +S}$ under charge conjugation 
and $P=(-1)^{\ell+1}$ under parity
(which gets broken when chiral gauge interactions of SM fermions play a role).

\medskip

The resulting 5-plet bound states at constituent mass $M=14\TeV$ 
are plotted in  fig.\fig{Levels14TeV}, and table~\ref{tab:MainLevels} lists their main properties.\footnote{While we agree with the generic formul\ae{}
for the rates in~\cite{1702.01141}, we found a missing order one factor in the application to the 5-plet:
the decay rates of the $2p$ states differ from eq.~(91) in~\cite{1702.01141} because a $\alpha_2$ should be $\alpha_{\rm eff}$.
These $2p$ decay rates negligibly affect the cosmological relic abundance computed in~\cite{1702.01141}.}
Bound states can decay via annihilation of their constituents with rate $\Gamma_{\rm ann}$,
or into deeper states with rate $\Gamma_{\rm dec}$.
Bound states with $J\equiv \ell \oplus S$ equal to 1 or 5 can annihilate into two $\SU(2)_L$ vectors $VV$,
while vector bound states with $J=3$ cannot because of the Landau-Yang theorem.

We are especially interested in bound states that annihilate into SM fermions, as they can
thereby be directly produced in $\mu^+\mu^-$ collisions.
Such states are those with the same quantum numbers as the weak vectors $W^a_\mu$,
so that such bound states mix and inherit couplings to fermions.
These special bound states are the $^n_1s_3^{--}$ vector triplets with $S=1$, $\ell=0$
(the full notation is explained in the first row of table~\ref{tab:MainLevels}), which decay into SM fermions with rate 
$\Gamma_{\text{ann},f}=625 M/2n^3 = 0.96\,\Gamma_{\text{ann}}$.
The other vector triplet, the $^2_1p_3^{+-}$ bound state with $S=0$, $\ell=1$,
has opposite parity and annihilates in $VVV$ rather than in fermions.
The bound state $^3_Jd_3^{--}$ with $\ell=2$ has the right quantum numbers,
but annihilation rates of bound states with $\ell >0$ are suppressed by extra powers of $\alpha_2$.

Table~\ref{tab:MainLevels} also shows the decay widths among bound states:
their computation will be discussed in section~\ref{boundcolldec}, where we discuss the associated collider signals.

\section{Bound-state production at colliders}\label{boundcoll}
All bound states have narrow total width, $\Gamma = \Gamma_{\rm ann} + \Gamma_{\rm dec}\ll M$.
Then, their collider phenomenology is well approximated
{\em à la} Breit-Wigner  such that their decay widths determine their production rates.
The cross section for $s$-channel production is
 \beq \sigma(i_1 i_2\to B \to f) \approx 
{\rm BW}(s) \sigma_{\rm peak}\eeq
where
\beq
{\rm BW}(s)  =  \frac{M_B^2 \Gamma_B^2}{(s-M_B^2)^2+M_B^2 \Gamma_B^2} \simeq \Gamma_B M_B {\pi}\,\delta(s-M_B^2),\qquad
 \sigma_{\rm peak}=\frac{16\pi S_B}{M_B^2 S_{i_1}S_{i_2} }  \hbox{BR}_{i_1 i_2} \hbox{BR}_f  
  \eeq 
and $S_i$ is the spin times group multiplicity of the various particles
(e.g.\ 2 for $\mu^\pm$, 3 if $B$ is a vector singlet etc).

The cross section needs to be convoluted with the energy distribution of a muon collider, described by some function $\wp(s)$ normalized as
$\int \wp(s) d\sqrt{s} =1$.
Assuming that each beam has a Gaussian energy distribution with standard deviation $\sigma_E$ one gets
a Gaussian distribution
\begin{equation}
\wp(s) = \frac{1}{\sqrt{2 \pi }\Delta_E} \exp \left[-\frac{(\sqrt{s}-M_B)^2}{2 \Delta_E^2}\right] \, , \qquad \Delta_E = \sqrt{2} {\sigma_E}.
\end{equation}
The energy resolution of a muon collider is expected to be
$\sigma_E \approx 10^{-3} E  \sim 14\GeV$~\cite{1901.06150}, larger than the widths of bound states,
$\Gamma_B \circa{<}\GeV$.
Thereby a muon collider cannot sit at the peak of the resonances, where the cross section
is as large as allowed by unitarity.
In the limit $\sigma_E \gg \Gamma_B$ the convoluted cross section is
 \beq \label{eq:sigmaconv}
 \sigma(i\to B \to f)
  \simeq \epsilon \sigma_{\rm peak},\qquad  \epsilon=\frac{\sqrt{\pi} \, \Gamma_B}{4\sigma_E}  . \eeq
 Thanks to the $\sigma_E$ at the denominator, bound states that
 can be directly produced from $i_1 i_2 = \mu^-\mu^+$ collisions
can have cross sections comparable or bigger than tree-level SM cross sections,
$\sigma \approx 4\pi \alpha_2^2/ s$.

\subsection{Production of $^ns_3$ bound states from $\mu^-\mu^+$ collisions}\label{ns3}
We here study states that can be directly produced from $\mu^-\mu^+$ collisions
with a resonant $s$-channel cross section.
These are the states with the same quantum numbers as electroweak vectors:
$I=3$, $S=1$ and $PC=--$, achieved in view of the constituent fermion 5-plets ${\cal X}$.
The first such state is $^{n=1}_{J=1}s_{I=3}^{--}$
($^1s_3$ for short), that exists for $M \circa{>}4.4\TeV$.
Table~\ref{tab:MainLevels} shows that, for $M=14\TeV$,  $^ns_3$ 
bound states exist for $n=\{1,2,3\}$.
The leading-order cross section for $s$-channel 
production of their neutral component $B^0$ is given by eq.\eq{sigmaconv} with
\beq \label{eq:xsec-analytic}
\epsilon \approx \frac{1}{192 n^3} \frac{10^{-3}}{\sigma_E/E},\qquad
\sigma_{\rm peak}(\mu^+  \mu^- \to B^0_{1s_3} \to f \bar f) = \frac{3 \pi}{M^2} \,\BR_{\mu}  \BR_{f} \approx 30\fb \frac{ \BR_{f} }{ \BR_{\ell}}
\eeq
 where $\BR_{\ell}= 1/25$ for any lepton flavour, and $\BR_{q}=3/25$ for any quark flavour.
 The denominator is 25 (rather than 24), taking into account the $1/25$ branching ratio into the Higgs multiplet.
A more precise evaluation includes higher order effects.
In particular, the signal cross section gets reduced by about a factor 2 taking into account initial state radiation
(ISR)  of $\gamma$ and $Z$.
We perform MonteCarlo simulations by  approximating such bound states as vectors $B_{n\mu}^a$
coupled as $ g_n B_{n\mu}^a (\bar f \gamma_\mu T^a f)_L$ to left-handed SM fermions,
and choosing couplings $g_n$ that reproduce the bound-state widths.
Then, numerical results from~{\sc  MadGraph}~\cite{MadGraph} show that $\gamma$ radiation dominates.
Such effect is analytically approximated by assigning a parton distribution function to each muon beam,
such that the amount of muons with energy equal to the beam energy gets reduced by
an order unity factor, analytically given by $\sim (\Gamma/M)^{4\alpha_{\rm em} \ln (E/m_\mu)/\pi}$~\cite{1607.03210}.
Precise analytical results~\cite{hep-ph/0006359} agree with numerical results.

 \medskip

  \begin{figure}[t]
\centering
$$\includegraphics[width=0.75\textwidth]{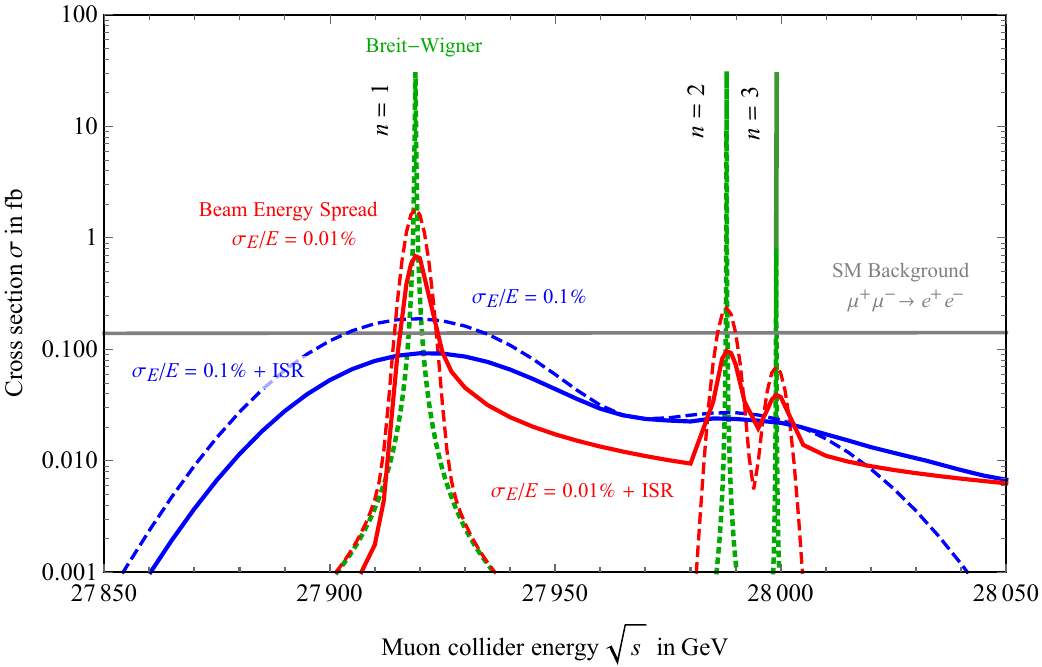}$$
\caption{\em\label{fig:peaks}
Bound-state signals of a Minimal Dark Matter 5-plet with constituent mass $M=14\TeV$.
The dotted green curves show the signal cross section for production of $^ns_1$ DM bound
states with $n=\{1,2,3\}$, ignoring the beam energy spread.
The dashed curves show the signal cross section, for two different values of the beam energy spread,
$\sigma_E = 10^{-3} E$ (baseline value) and $\sigma_E = 10^{-4} E$ (feasible value).
The continuous curves show the signal cross section after also taking into account initial state emission.
The gray horizontal curve is the SM $\mu^+\mu^- \to e^+ e^-$ background.}
\end{figure}

 \begin{figure}[t]
\centering
$$\includegraphics[width=0.335\textwidth]{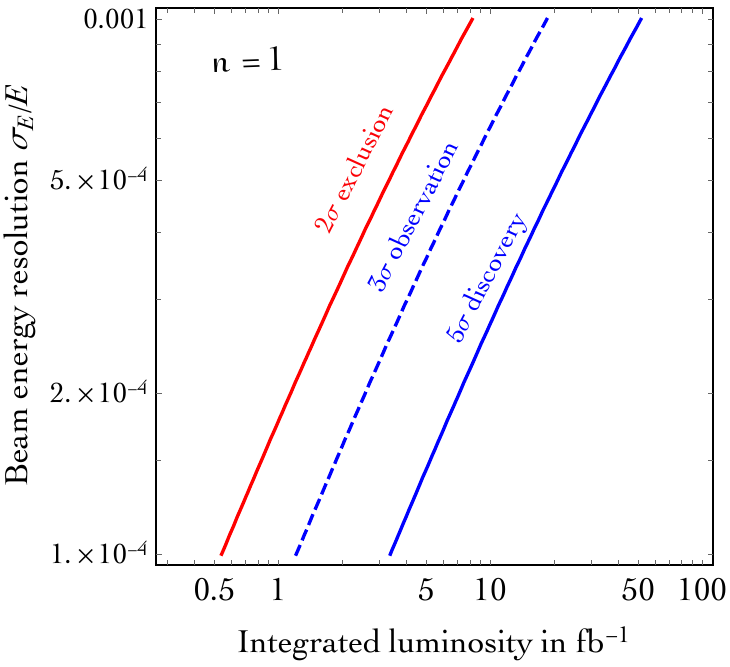}\quad
\includegraphics[width=0.31\textwidth]{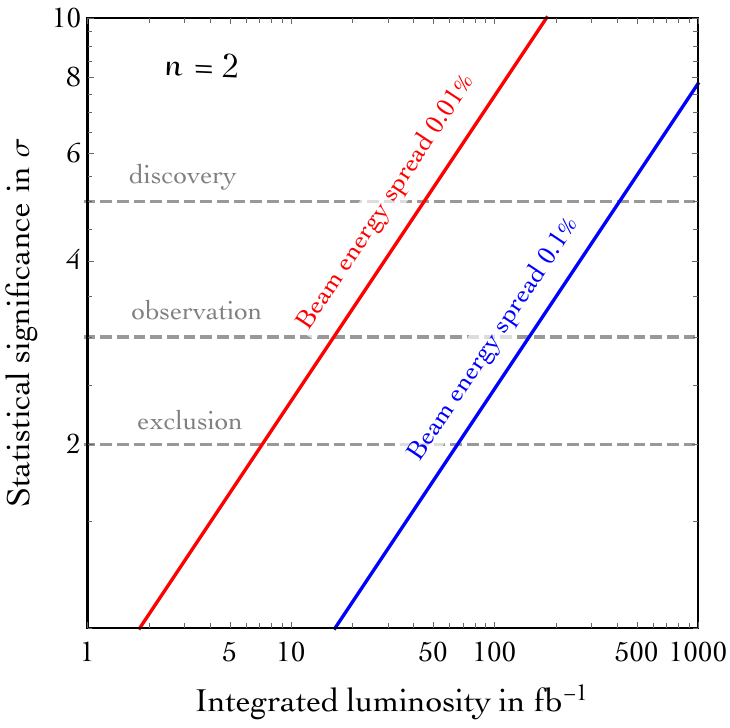}\quad
\includegraphics[width=0.31\textwidth]{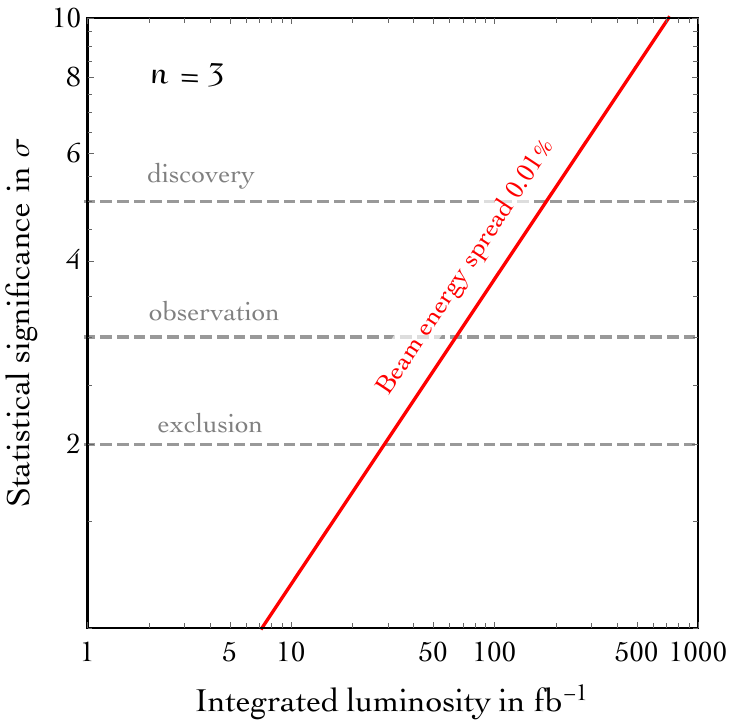}$$
\caption{\em\label{fig:reach} The integrated luminosity needed to see the $^n{}s_3$ bound states 
with mass $2M=28\TeV$ of a Minimal DM 5-plet at a muon collider with $s\approx 4M^2$
is much smaller than the possible value, $90/\ab $ at $\sigma_E/E = 10^{-3}$.}
\end{figure}

Considering, for example, the $e^-e^+$ final state (so that calorimeters can precisely measure their large energy), the SM background is
 \beq \label{eq:ee->mumu}
\sigma_{\rm SM}(\mu^+\mu^- \to e^+ e^- ) = 
\frac{4\pi \alpha^2_{\rm em}}{3s} +\frac{2\pi \alpha_{\rm em}\alpha_2}{3 \cW^2 s}(g_{L}+g_{R})^2  +\frac{\pi  \alpha_2^2}{3 \cW^4 s}
(g_{L}^2+g_{R}^2)^2  \approx 140\ab  \frac{(28\TeV)^2}{s}
\eeq
where $s\gg M_Z^2$,  $\cW=M_W/M_Z$, $g_L=1/2-\cW^2$, $g_R=1-\cW^2$.
We see that $\sigma_{\rm peak}$ is 200 times larger than $\sigma_{\rm SM}$
(green dotted curve in fig.\fig{peaks})
and that a design energy spread reduces it by $\epsilon \sim 1/200$ for $n=1$,
providing a DM signal at the level of total SM backgrounds
(dashed blue curve in fig.\fig{peaks}). The $n=1$ state can be mildly separated from those with $n=\{2,3\}$, 
that have rates below the SM background and thereby need some dedicated search.

Fig.\fig{reach}a shows that the integrated luminosity needed to discover such state corresponds to about one day of running,
taking into account its annihilation channels into $e^+ e^-$ and jets and without performing selection cuts
(for simplicity, we do not include annihilations into $\mu^+\mu^-$, which have a larger background due to $t$-channel vector exchange 
that can be
efficiently reduced by cuts on $p_T$ and other variables).
We assumed a $70\%$ efficiency for detecting each electron or jet in the final state.
Reducing the beam energy spread reduces the needed integrated luminosity,
but by an amount similar to the expected loss in collider luminosity. 

With a feasible reduction of $\sigma_E$ by one order of magnitude,
the cross section for producing the $^1s_3$ bound state becomes one order of magnitude larger than
SM backgrounds, and the excited bound states with $n=2,3$ can be separated and acquire total cross sections at the level of the 
SM backgrounds (dashed red curve in fig.\fig{peaks}).
 After taking initial state radiation into account, one obtains the continuous curves in fig.\fig{peaks}, where
peaks become asymmetric and larger above the threshold due to the  `radiative return' phenomenon. 
Fig.\fig{reach}b,c show the integrated luminosity needed to discover such states. 
The non-resonant loop corrections considered by~\cite{Katayose} 
at $\sqrt{s}$ slightly above the $2M$ threshold interfere destructively with the SM background 
leading to a decrease of the SM cross section by up to $8\%$.

 \medskip
 
\begin{figure}
\begin{center}
$$\hspace{5ex}\includegraphics[width=0.47\textwidth,angle=0]{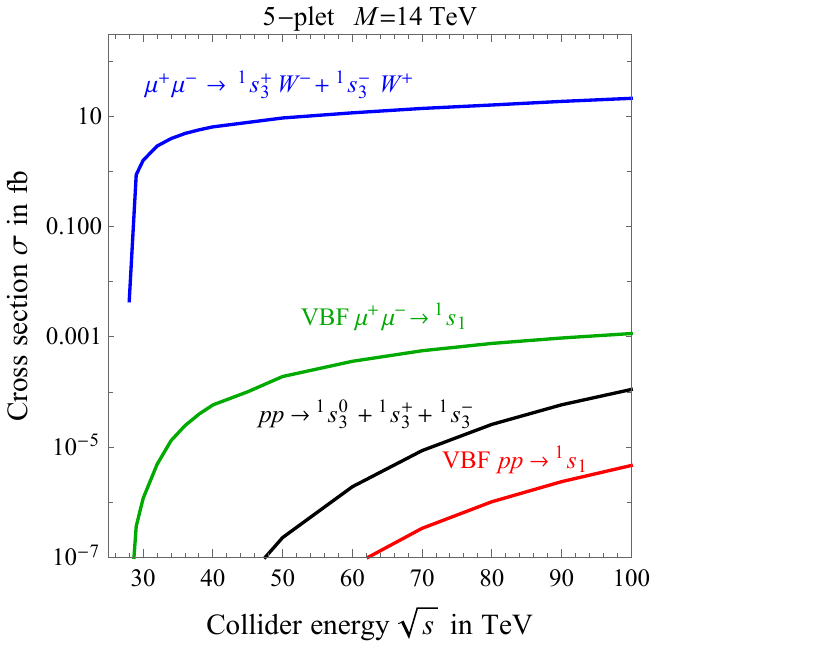}
\includegraphics[width=0.47\textwidth,angle=0]{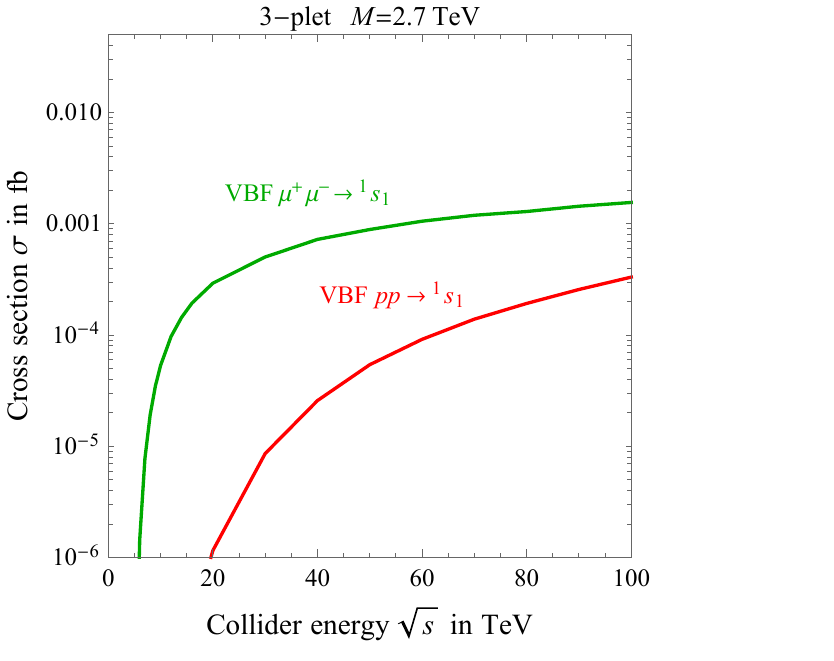}$$
\caption[]{\em\label{fig:evens} \small Cross sections for the production of 
some bound states of the Minimal DM fermionic 5-plet with constituent mass
$M=14\TeV$ (left)
and of the fermionic 3-plet with $M=2.7\TeV$ (right)
at a $\mu^+\mu^-$ collider and at a $pp$ collider.  }
\end{center}
\end{figure}

The charged components $B^\pm_\mu$ of the isospin triplet of bound states are produced with a relatively large cross
section, given that the partonic neutrino component of a $\mu^\pm$ beam is
peaked at energy fraction
$x=1$~\cite{2007.14300}, in view of soft $W^\pm$ emission.
By running  a bit above the peak, the state with $n=1$ is produced 
as $\mu^+ \mu^- \to B_\mu^\pm W^\mp_\mu$ with fb-scale cross section, as shown by the blue curve in fig.\fig{evens}.

Finally, we mention that the state $^3d_3$ too has the same quantum numbers as electroweak vectors
and can thereby be produced directly from $\mu^+\mu^-$ collision;
however its annihilation rate (see bottom row of table~\ref{tab:MainLevels})
is highly suppressed by $\alpha_2^{5+2\ell}$ in view of $\ell=2$ 
and we neglect it.

\subsection{Production of other bound states from $VV$ collisions}\label{boundcollnonres}
The other bound states annihilate to weak vectors and can thereby
be produced through associated production via vectors.
Then, the  energy spread in the effective collision energy becomes large, $\sigma_E \sim \sqrt{s}$,
and the cross sections small. 
These more general processes can be computed using automated codes~\cite{MadGraph},
approximating bound states as particles
with effective couplings to their decay products that reproduce the widths~\cite{0912.0526}
computed in table~\ref{tab:MainLevels}.
For example, the ground pseudo-scalar bound state $^1s_1$ with $I=1$
can be written as a scalar singlet $B$ coupled as
$B \epsilon_{\mu\nu\mu'\nu'}V^a_{\mu\nu} V^a_{\mu'\nu'}$.
The pseudo-scalar bound state $^1s_5$  with $I=5$ can be written as a scalar 
 $B_{aa'}$ in the symmetric trace-less representation of $\SU(2)_L$   coupled as
$B_{aa'} \epsilon_{\mu\nu\mu'\nu'}V^a_{\mu\nu} V^{a'}_{\mu'\nu'}$.

\smallskip

We focus on the ground state $^1s_1$, as it has the largest annihilation rate.
It can be produced
via scatterings of SM electroweak vectors, $\gamma \gamma \to B_{{}1s_1}$,  $\gamma \mu^\pm \to B_{{}1s_1}  \mu^\pm$, $\mu^+ \mu^- \to B_{{}1s_1}  \mu^+ \mu^-$, $\mu^+ \mu^- \to B_{{}1s_1}  \nu_\mu \bar{\nu}_\mu$.
Resonant production is not possible and one thereby must run at higher $\sqrt{s}> 2M$: 
the green curve in fig.\fig{evens} shows that, as expected, the production cross section is much smaller.

The red curve in fig.\fig{evens} shows its analogous production cross sections at a $pp$ collider, 
which is even smaller given that vector partons in a
$p$ beam have lower energy than in a $\mu$ beam. 
For completeness, the black curve in fig.\fig{evens} shows the cross section for production at a $pp$ collider
of the $^1s_3$ bound state discussed in the previous section.
We do not discuss the backgrounds.

\medskip

Furthermore, we consider a Wino-like Minimal DM fermionic triplet.
The DM abundance is reproduced thermally for $M=2.7\TeV$.
At this mass only one $^1s_1$ bound state exists with $E_B\approx 68\MeV$ and $\Gamma_B = 8 \alpha_2^5 M \approx 4\MeV$~\cite{1702.01141}.  
This bound state cannot be produced with a resonantly-enhanced cross section.
Fig.\fig{evens}b shows its production cross section at a muon or $pp$ collider.

\subsection{Decays of bound states and their collider signals}\label{boundcolldec}
In this section we describe the computation of the bound state decays
listed in table~\ref{tab:MainLevels}, having in mind that we seek characteristic 
collider signals produced by decays among bound states.
The leading-order decays $B\to B' V$
proceed through the emission of a weak vector boson $V$,
which is often a photon as the phase space for $W,Z$ emission is often closed.
Such process dominantly occurs via electric dipole transitions,
although magnetic dipole transitions happen to be important in cases
where selection rules forbid electric dipole transitions.
We compute bound states in the $\SU(2)_L$-symmetric approximation,
so that bound states of two 5-plets have
two indices $ij$ in the 5 representation,
that can be converted into isospin eigenstates $B_{\tilde{I}\tilde{I}_3}$ through Clebsch-Gordan coefficients:
$ B_{ij}  = C_{ij}^{\tilde{I}\tilde{I}_3} B_{\tilde{I}\tilde{I}_3}$.
\begin{itemize}
\item The effective interaction hamiltonian for the electric dipole at leading order is
\begin{equation}
\label{eq:ed}
  H_{\rm el} = -\frac{g_2}{M}[\vec{A}^a(x_1)\cdot \vec{p}_1\,  T^a_{i'i}\delta_{jj'}+
  \vec{A}^a(x_2)\cdot \vec{p}_2\, \bar{T}^a_{j'j}\delta_{ii'}]+
   g_2\alpha_2 [\vec{A}^a(0)\cdot \hat{r}]T^b_{i'i}\bar{T}^c_{j'j}f^{abc}
\end{equation}
leading to the following selection rules:
$|\Delta \tilde I|=1$, $|\Delta \ell|=1$, $\Delta S=0$.
Decay rates are obtained as
{\small\begin{equation}
\Gamma(^2p_{\tilde{I}}\rightarrow {}^1s_{\tilde{I}'}+V^a)=\frac{16}{9I_{2p}}\frac{\alpha_2k}{M^2}
\sum_{\tilde{I}_3\tilde{I}'_3}\left|\int r^2\mathrm{d}r \, R_{\tilde{I},2p}\left(C_\mathcal{J}^{a\tilde{I}_3\tilde{I}'_3}\partial_r-C_\mathcal{T}^{a\tilde{I}_3\tilde{I}'_3}\frac{\alpha_2 M}{2}\right)R_{\tilde{I}',1s}\right|^2
\end{equation} 
\begin{equation}
\Gamma(^3s_{\tilde{I}}\rightarrow {}^2p_{\tilde{I}'}+V^a)=\frac{16}{3I_{3s}}\frac{\alpha_2k}{M^2}
\sum_{\tilde{I}_3\tilde{I}'_3}\left|\int r^2\mathrm{d}r \,R_{\tilde{I}',2p}\left(C_\mathcal{J}^{a\tilde{I}_3\tilde{I}'_3}\partial_r+C_\mathcal{T}^{a\tilde{I}_3\tilde{I}'_3}\frac{\alpha_2 M}{2}\right)R_{\tilde{I},3s}\right|^2
\end{equation}}
$\!\!$where $r$ is the radius,
$R(r)$ are normalized radial wave-functions, $k$ is the spatial momentum of $V$,
$I$ is the isospin of the initial bound state, and
\begin{equation}
C_\mathcal{J}^{a\tilde{I}_3\tilde{I}'_3}=\frac{1}{2}\mathrm{Tr}\left[C^{\tilde{I}' \tilde{I}'_3}\left\{C^{\tilde{I} \tilde{I}_3},T^a\right\}\right],\qquad
C_\mathcal{T}^{a\tilde{I}_3\tilde{I}'_3}=i\mathrm{Tr}\left[C^{\tilde{I}' \tilde{I}'_3}T^b C^{\tilde{I} \tilde{I}_3}T^c\right]f^{abc}.
\end{equation}

\item The effective interaction hamiltonian for the magnetic dipole at leading order is
(see e.g.~\cite{2007.07231})
\begin{equation}\label{eq:Hmag}
 H_{\rm mag} = -\frac{g_2}{2M}[T^a_{i'i}\delta_{jj'}\vec{\sigma}\cdot \vec{{B}}^a(x_1)+\bar{T}^a_{j'j}\delta_{ii'}\vec{\sigma}\cdot \vec{{B}}^a(x_2)] +\cdots
\end{equation}
leading to the following selection rules:
$|\Delta \tilde I|=1$, $\Delta \ell=0$, $|\Delta S|=1$.
Decay rates are obtained as~\cite{2007.07231}
\begin{equation}
\begin{split}
\Gamma(^{n_i}s_{\tilde{I}_i}\rightarrow {}^{n_f}s_{\tilde{I}_f}+V^a)&=\frac{2^3}{I_i}\frac{\alpha_2k^3}{M^2}\sum_{\tilde{I}_{3,i}\tilde{I}_{3,f}}\left|C_{\mathcal{J}}^{a\tilde{I}_{3,i}\tilde{I}_{3,f}}
\int r^2\mathrm{d}r \, R_{n_is_{\tilde{I}_i}}R_{n_fs_{\tilde{I}_f}}\right|^2
\end{split}
\end{equation}
with no contribution from the omitted 
non-abelian term in eq.\eq{Hmag}.

\item Higher-order interactions lead to multiple-vector emission, with suppressed rates
that turn out to be negligible.

\end{itemize}
As discussed in section~\ref{ns3},
the lightest bound state that can be produced resonantly is the neutral component of $^1s_3$. 
This is the only component of $^1s_3$ that can decay
($W^\pm$ emission from charged components of $^1s_3$ is kinematically blocked)
to $^1s_1 \gamma$ via a magnetic transition
with a rate $\Gamma_{\rm dec} = 3 \times 4.6\keV$.
Such rate is of order $\alpha_2^6 \alpha_{\rm em} M$,
 where an $\alpha_2^2$ factor arises from the $\gamma$ phase space;
another $\alpha_2^4$ from the magnetic field $\vec{B}^a$;
the $ \alpha_{\rm em} $ from photon emission.
Taking into account its annihilation rate, the neutral component of the
$^1s_3$ bound state 
decays into  a monochromatic $\gamma$ with energy $E_\gamma \approx 38 \GeV$
with branching ratio  $\hbox{BR}_{\rm dec} \approx 9~10^{-5}$.
This corresponds to 19 events in a run with baseline $\sigma_E=10^{-3}$ and luminosity
${\cal L}=90/\ab$.

\medskip

Higher order states are produced with a lower cross section, that scales as $\Gamma_{\rm ann} \propto 1/n^3$.
Nevertheless, such states could give a higher rate of decay events, proportional to
$\sigma ~{\rm BR}_{\rm dec} \propto \Gamma_{\rm ann} \times  \Gamma_{\rm dec}/ \Gamma_{\rm ann} 
\propto  \Gamma_{\rm dec} $. 

\begin{itemize}
\item  At $n=2$, the $^2s_3$ bound state similarly decays magnetically, with the difference
that it can now also emit massive weak bosons, and decay into multiple states $^1s_1$, $^1s_5$
(we neglect decays in $^2s_1$ because their rate is negligibly small, at eV level).
The neutral component of $^2s_3$ decays emitting a $\gamma$ with rate
$\Gamma_{\rm dec}\approx 2.0\keV$ and emitting a $Z$ with rate
$\Gamma_{\rm dec}\approx 1.7 \keV$;
charged components have similar decay rates.
In view of  the lower binding energy and wave-function overlap,
$^2s_3$ thereby gives a similar number of decay events as $^1s_3$.
As a result, the $\gamma$ decays of the  $^2s_3$ neutral component
produces two distinctive single-photon lines,
at $E_\gamma \approx 105\GeV$ and $13\GeV$,
as well as $Z$ bosons.


\item At $n=3$, the $^3s_3$ bound state can decay electrically into $^2p_{1+5}\gamma$,
with a rate of order $\alpha_2^4 \alpha_{\rm em } M$
(where an $\alpha_2^2$ factors arises from the $\gamma$ phase space;
another $\alpha_2^2$ from the dipole matrix element;
the $ \alpha_{\rm em} $ from photon emission). 
The numerical coefficient turns however to be small,
and the decay rate is again around a keV.
More precisely, only the neutral component can decay into $^2p_1$,
and all components decay equally into $^2p_5$.
Thereby, table~\ref{tab:MainLevels} implies that the decay rate of the neutral component is
$\Gamma_{\rm dec}= (3\times0.003 + 0.005)\alpha_2^4\alpha_{\rm em}\approx 1.7\keV$.
The $^2p_{1+5}$ bound states next dominantly decay via a large electric dipole
into $^1s_3$, that annihilates. 
This process thereby gives a set of multiple-photon lines, with
$E_\gamma \approx \{18\GeV, 60\GeV\}$, and with
$E_\gamma\approx \{0.5\GeV, 79\GeV\}$.
As signal events have very distinctive $\gamma\gamma$ signatures, 
backgrounds can be strongly reduced.

\end{itemize}
For completeness, in table~\ref{tab:MainLevels}
we also computed decay rates of other states that cannot be produced resonantly with large rates.
Thereby we do not discuss them.
All above numbers assume $M=14\TeV$ and need to be recomputed otherwise.

\section{Conclusions}\label{concl}
We considered DM as electroweak multiplets,
and studied the effects of their electroweak bound states at future colliders.
We found that bound states with the same quantum numbers as electroweak vectors
can be produced resonantly with large cross sections by running lepton colliders at the appropriate $\sqrt{s}$.
Such bound states exist if DM is a heavy enough fermionic multiplet.\footnote{A similar resonant enhancement arises from bound states with the same quantum numbers as the Higgs boson,
possibly present in speculative DM models where the Higgs 
mediates attractive forces between DM constituents, fermionic or bosonic.}
A wino-like weak fermionic triplet with `thermal' mass that reproduces the cosmological DM density,
$M=2.7\TeV$, does not form such bound states.
Three of such bound states 
arise if DM is an automatically-stable 
fermionic weak 5-plet with `thermal' mass $M \approx 14\TeV$.
The level structure is plotted in fig.\fig{Levels14TeV}, the main properties of the bound states are computed in
table~\ref{tab:MainLevels}, and
fig.\fig{peaks} shows that the three predicted peaks would be easily observable at a muon collider,
running at $\sqrt{s}\approx 2M$ and with a beam energy spread $\sigma_E/E = 10^{-3}$ or better.
The production rates of the neutral components $B^0_n$ of the bound-state triplets
are so large that one day of running may be enough for discovery, see fig.\fig{reach}.
One could next search for rarer but more characteristic sets of single and multiple lines  $\gamma$ 
produced by decays among bound states, as discussed in section~\ref{boundcolldec}.
The charged components $B^\pm_n$ of the bound state triplets
can also be produced with partially enhanced cross section
at a broader $\sqrt{s}\circa{>} 2M$,
as neutrinos have a peaked partonic distribution in muons.

We also studied the other bound states that cannot be produced resonantly.
The extra 5-plet bound states in table~\ref{tab:MainLevels},
as well as the only bound state formed by a 3-plet,
are produced at $\mu^+\mu^-$ or $pp$ colliders 
with small non-resonant cross sections as shown in fig.\fig{evens}.

\medskip

As an aside final comment, we mention a new Minimal DM signal even more
futuristic than a muon collider.\footnote{A.S.\ thanks Paolo Panci and 
Raghuveer Garani for pointing it out.}
DM ${\cal X}^0$ gravitationally attracted by a neutron star reaches relativistic velocity before hitting its surface,
so that charged current scatterings such as ${\cal X}^0 n \to {\cal X}^- p$
and ${\cal X}^0 p \to {\cal X}^+ n$ become kinematically allowed despite the $\Delta M \approx 166\MeV$ gap,
and have a large tree-level cross section $\sigma \sim m_n^2/v^4 \sim 10^{-38}\cm^2$.
Decays of charged components can then produce neutrinos with energy around $10\MeV$ at 
rate $\dot N_\nu \sim \dot M/\Delta M \sim 10^{25}/{\rm sec}$.
Here $\dot M_{\rm DM} = \rho_{\rm DM} v_{\rm DM} \pi b^2$ is the DM mass
that falls in the neutron star.
The impact parameter is $b/R_{\rm ns} = v_{\rm esc}/v_{\rm DM}/ \sqrt{1-v_{\rm esc}^2 }$
where $R_{\rm ns} \sim (M_{\rm Pl}/m_n)^3 \sim \hbox{few km}$ is the radius of the neutron star,
and $v_{\rm esc} \sim 1$ is the escape velocity.
Given that the nearest neutron stars are expected to be at distance $d \sim {\rm pc}$ from the Earth,
the resulting neutrino flux $\Phi_\nu \sim \dot N_\nu/4\pi d^2$ is about 10 orders of magnitude
below the expected background of supernova neutrinos.
A very futuristic detector close to a neutron star is needed to reveal the signal.

\small

\paragraph{Acknowledgements}
We thank Michele Redi, Andrea Wulzer and Ivano Basile.
This work was supported by the ERC grant 669668 NEO-NAT and by PRIN 2017FMJFMW.

\end{document}